\documentstyle[11pt,aaspp4,flushrt,tighten]{article}  % preprin 1 colonna
\lefthead{authors }
\righthead {title}

\def\msun{{M_\odot}}
\def\nHI{{\rm HI}}
\def\nH{{\rm H}}

\def\kmsmpc{\,{\rm km\,s^{-1}\,Mpc^{-1}}}
\def\HI{\hbox{H~$\scriptstyle\rm I\ $}}
\def\bHI{\hbox{\bf H~$\scriptstyle\bf I\ $}}
\def\HII{\hbox{H~$\scriptstyle\rm II\ $}}

\def\spose#1{\hbox to 0pt{#1\hss}}
\def\lta{\mathrel{\spose{\lower 3pt\hbox{$\mathchar"218$}}
     \raise 2.0pt\hbox{$\mathchar"13C$}}}
\def\gta{\mathrel{\spose{\lower 3pt\hbox{$\mathchar"218$}}
     \raise 2.0pt\hbox{$\mathchar"13E$}}}
\def\etal{{et al.~}}
\def\Lya{Ly$\alpha\ $}
\begin{document}

\title{Radio Signatures of \bHI at High Redshift:\ Mapping the End of 
the ``Dark Ages''}

\author{Paolo Tozzi\altaffilmark{1,2}, Piero Madau\altaffilmark{1,3}, 
Avery Meiksin\altaffilmark{4}, and Martin J. Rees\altaffilmark{3}} 

\altaffiltext{1}{Space Telescope Science Institute, 3700 San Martin Drive,
Baltimore MD 21218.}  
\altaffiltext{2}{Department of Physics and Astronomy, The Johns Hopkins 
University, Baltimore, MD 21218.} 
\altaffiltext{3}{Institute of Astronomy, Madingley Road, Cambridge CB3 0HA, 
UK.} 
\altaffiltext{4}{Institute for Astronomy, University of Edinburgh, Royal 
Observatory, Edinburgh EH9 3HJ, UK.}

\begin{abstract}
\noindent The emission of 21--cm radiation from 
a neutral intergalactic medium (IGM) 
at high redshift is discussed in connection with the thermal and ionization 
history of the universe. The physical mechanisms that make 
such radiation detectable against the cosmic microwave background include 
Ly$\alpha$ coupling of the hydrogen spin temperature to the kinetic 
temperature of the gas and preheating of the IGM
by the first generation of stars and quasars. Three different signatures are 
investigated in detail: ({\it a}) the fluctuations in the redshifted 
21--cm emission induced by the gas density inhomogeneities that develop 
at early times in cold dark matter (CDM) dominated cosmologies; ({\it b}) 
the sharp absorption feature in the radio sky due to the rapid rise of the 
Ly$\alpha$ continuum background that marks the 
birth of the first UV sources in the universe; and ({\it c}) the 21--cm 
emission and absorption shells that are generated on several Mpc scales 
around the first bright quasars. Future radio observations with 
projected facilities like the  
{\it Giant Metrewave Radio Telescope} and the {\it Square Kilometer Array} 
may shed light on the power spectrum of density fluctuations at $z>5$,
and map the end of the ``dark ages", i.e. the transition from 
the post--recombination universe to one populated with radiation sources.

\end{abstract}


\section{Introduction}

The development of structure in the universe was well advanced at high
redshifts. Quasars have been detected nearly to $z=5$, and the most
distant galaxies to even greater distances (Weymann \etal 1998). The
\Lya forest in the spectra of high redshift QSOs shows that the
intergalactic medium (IGM) itself developed extensive nonlinear
structures by early times. Still unknown, however, is the epoch during
which the first generation of stars was formed, and the time scale
over which the transition from a neutral universe to one that is
almost completely reionized took place. While numerical simulations
have shown that the IGM is expected to fragment into structures at
early times in cold dark matter (CDM) cosmogonies (Cen \etal 1994; Zhang 
\etal 1995; Hernquist \etal 1996; Zhang \etal 1998), 
the same simulations are much less able to predict the efficiency with
which the first gravitationally collapsed objects lit up the universe
at the end of the ``dark age'' (Rees 1998). The nature of the first
dominant ionizing sources is still unknown, and although IR
observations will permit even higher redshift galaxies to be
observed, detections become increasingly difficult because of the
surface brightness dimming from cosmological expansion.

Madau, Meiksin, \& Rees (1997, hereafter MMR) demonstrated that radiation 
sources prior to the epoch of reionization may be detected indirectly
through their impact on the surrounding neutral IGM and the resulting
emission or absorption against the cosmic microwave background (CMB)
at the frequency corresponding to the redshifted 21--cm line. 
While its energy density is estimated to be about two orders of 
magnitude lower than the CMB, the 21--cm spectral signature will display 
angular structure as well as structure in redshift space due to
inhomogeneities in the gas density field (Hogan \& Rees 1979; Scott 
\& Rees 1990), hydrogen ionized fraction, and spin temperature (MMR).
Because of the smoothness of the CMB sky, fluctuations in the 21--cm
radiation will dominate the CMB fluctuations by about 2 orders of magnitude
on several arcmin scales.
Radio observations at meter wavelengths have therefore the potential to 
tell us how (as well as when) the primordial gas was reheated and reionized:
if the reheating was due to ultraluminous but sparsely distributed sources 
like QSOs, or by fainter but more uniformly distributed objects like an 
abundant population of  pregalactic stars.

In this paper we discuss possible ways in which the next generation of
radio telescopes could provide the first observations of a neutral IGM
at a level of a few tenths of $\mu$Jy arcmin$^{-2}$. The proposed {\it
Square Kilometer Array} ({\it SKA}) would provide an enhancement of
about two orders of magnitude in sensitivity relative to present--day
radio telescopes (Taylor \& Braun 1999). We discuss the technical requirements
necessary to detect the 21--cm signature from the IGM and describe a
possible design for a special purpose telescope. We emphasize that within
existing technological limits, radio astronomy could effectively open up much
of the universe to a direct study of the reheating and reionization epochs.

\section{Basic theory}

The ground state hyperfine levels of hydrogen will very quickly reach
thermal equilibrium with the CMB. In order to detect the IGM in either
emission or absorption, it is crucial that an effective mechanism be
present to decouple the hydrogen spin states from the CMB. Following
MMR, we briefly review below the mechanisms that determine the level
populations of the hyperfine state of neutral hydrogen in the
intergalactic medium (see also Meiksin 1999).

\subsection{Spin temperature}

The emission or absorption of 21--cm radiation from a neutral IGM is
governed by the hydrogen spin temperature $T_S$, defined by
\begin{equation}
\frac{n_1}{n_0}=3\exp(-T_*/ T_S),
\end{equation}
where $n_0$ and $n_1$ are the singlet and triplet $n=1$ hyperfine
levels, $T_*\equiv h_P\nu_{10}/k_B=0.07\,$K, $\nu_{10}$ is the
frequency of the 21--cm transition, $h_P$ is Planck's constant, and $k_B$
is Boltzmann's constant. In the presence of only the CMB radiation with 
$T_{\rm CMB}=2.73\,(1+z)\,$K, the spin states will reach thermal equilibrium
with the CMB on a timescale of $T_*/(T_{\rm CMB}A_{10})\approx 3\times10^5\,
(1+z)^{-1}\,$yr ($A_{10}=2.9\times 10^{-15}\,$s$^{-1}$
is the spontaneous decay rate of the hyperfine transition of atomic hydrogen),
and intergalactic \HI will produce neither an absorption nor an emission 
signature. A mechanism is required that decouples $T_S$ and $T_{\rm CMB}$, 
e.g. by coupling the spin temperature instead to the kinetic temperature
$T_K$ of the gas itself. Two mechanisms are
available, collisions between hydrogen atoms (Purcell \& Field 1956)
and scattering by \Lya photons (Wouthuysen 1952; Field 1958). The
collision--induced coupling between the spin and kinetic temperatures
is dominated by the spin--exchange process between the colliding
hydrogen atoms. The rate, however, is too small for realistic IGM
densities at the redshifts of interest, although collisions may be
important in dense regions, $\delta\rho/\rho\gta 30[(1+z)/10]^{-2}$ (MMR).

In the low density IGM, the dominant mechanism is the scattering of continuum
UV photons redshifted by the Hubble expansion into local \Lya photons. The
many scatterings mix the hyperfine levels of neutral hydrogen in its
ground state via intermediate transitions to the $2p$ state, the
Wouthuysen--Field process. An atom initially in the $n=1$ singlet state absorbs
a \Lya photon that puts it in an $n=2$ state, allowing it to return to the
triplet $n=1$ state by a spontaneous decay. As the neutral IGM is highly 
opaque to resonant scattering, the shape of the continuum radiation spectrum 
near \Lya will follow a Boltzmann distribution with a temperature given by 
the kinetic temperature of the IGM (Field 1959). In this case the spin 
temperature of neutral hydrogen is a weighted mean between the matter and 
CMB temperatures, \footnote{In the presence of a radio source,
the antenna temperature of the radio emission should be added to $T_{\rm CMB}$
in eq. (\ref{eq:Tspin}). The radio emission may make an important
contribution in the vicinity of a radio--loud quasar (Bahcall \& Ekers 1969),
and would itself permit the IGM to be detected in 21--cm radiation. This may
be an important process if the emission of ionizing radiation is more tightly
beamed than the radio continuum emission, allowing the gas to remain neutral
near the quasar after it has switched on (MMR).}
\begin{equation}
T_S=\frac{T_{\rm CMB}+y_\alpha T_K}{1+y_\alpha},  \label{eq:Tspin}
\end{equation}
where
\begin{equation}
y_\alpha\equiv\frac{P_{10} T_*}{A_{10} T_K}
\end{equation}
is the \Lya pumping efficiency. Here $P_{10}$ is the indirect de--excitation 
rate of the triplet $n=1$ state via absorption of a \Lya photon to the $n=2$
level, and $T_S\gg T_*$ is assumed. If $y_\alpha$ is large, $T_S\rightarrow
T_K$, signifying equilibrium with the matter. A consideration
of the net transition rates between the various hyperfine $n=1$ and $n=2$
levels above shows that the $1\rightarrow0$ transition rate via \Lya scattering
is related to the total rate $P_\alpha$ by $P_{10}=4P_\alpha/27$ (Field 1958).
This relation and equation~(\ref{eq:Tspin}) are derived in the Appendix.
In the limit $T_K\gg T_{\rm CMB}$, the fractional deviation in a steady
state of the spin temperature from the temperature of the CMB is
\begin{equation}
{{T_S-T_{\rm CMB}}\over T_S}\approx \left(1+{T_{\rm CMB}\over y_\alpha
T_K}\right)^{-1}. \label{eq:ftspin}
\end{equation}
There exists then a critical value of $P_\alpha$ which, if greatly exceeded,
would drive $T_S\rightarrow T_K$. This ``thermalization'' rate is (MMR)
\begin{equation}
P_{\rm th}\equiv {27A_{10}T_{\rm CMB}\over 4T_*}\approx 
(7.6\times 10^{-12} ~{\rm s}^{-1})~ \left({1+z\over 10}\right).
\label{eq:ptherm}
\end{equation}

\subsection{Brightness temperature}

Consider a patch of neutral hydrogen with an angular diameter on the sky
larger than the beamwidth of the telescope, and a radial velocity width
broader than the bandwidth. In an Einstein--de Sitter universe with $H_0=100\, 
h\,\kmsmpc$ and baryon density parameter $\Omega_b$, the optical depth of the 
patch at $21 (1+z)\,$cm for a spin temperature $T_S\neq T_{\rm CMB}$ is
\begin{equation}
\tau(z)=\frac{3}{32\pi}\lambda_{10}^3A_{10}\frac{T_*}{T_S}\frac{n_\nHI(0)}{H_0}
(1+z)^{1.5}\approx 0.0033\,h^{-1}\left({T_{\rm CMB}\over T_S}\right)
\left({\Omega_bh^2 \over 0.02}\right) \left({1+z\over 10}
\right)^{1/2}.
\label{eq:tau}
\end{equation}
The optical depth will typically be much less than unity. Since the brightness
temperature through the IGM is
$T_b=T_{\rm CMB} e^{-\tau}+T_S(1-e^{-\tau})$, the differential antenna
temperature observed at Earth between this region and the CMB will be
\begin{equation}
\delta T_b=(1+z)^{-1} (T_S-T_{\rm CMB}) (1-e^{-\tau})\approx (9.0~{\rm mK})
h^{-1} \left({\Omega_bh^2\over 0.02}\right) 
\left({{1+z}\over 10}\right)^{1/2} \left({T_S-T_{\rm CMB}\over T_S}\right).
\label{eq:dT}
\end{equation}
If the intergalactic gas has been significantly preheated, $T_S$ will be much
larger than $T_{\rm CMB}$, and the IGM will be observed in
emission at a level that is independent of the exact value of $T_S$. By
contrast, when $T_{\rm CMB}\gg T_S$ (negligible preheating), the gas will
appear in absorption, and $\vert\delta T_b\vert$ will be a factor
$\sim T_{\rm CMB}/T_S$ larger than in emission, so that it becomes relatively
easier to detect intergalactic \HI (Scott \& Rees 1990). 

The presence of a sufficient flux of \Lya photons will thus render the neutral
IGM ``visible.'' Without heating sources, the adiabatic expansion of the
universe will lower the kinetic temperature of the IGM well below that of the
CMB, and the IGM will be detectable through its absorption. If there are
sources of radiation that preheat the IGM, it may instead be possible 
to detect the IGM in emission.

\subsection{Preheating}

The energetic demand for heating the IGM above the CMB temperature is meager,
only $\sim 0.004\,$ eV per particle at $z\sim 10$. Consequently, even
relatively inefficient heating mechanisms may be important warming sources
well before the IGM has been reionized.\footnote{In an inhomogeneous universe 
reheating appears to be a slow process occurring at early epochs, followed 
by sudden reionization at later times (Gnedin \& Ostriker 1997).}~
Possible preheating sources are soft X--rays from an early generation of
quasars or thermal bremsstrahlung emission produced by the ionized gas in
the collapsed halos of young galaxies (MMR).
While photons emitted by a quasar just shortward of the photoelectric edge 
are absorbed at the ionization front, photons of much shorter wavelength will
be able to propagate into the neutral IGM much further. Most of
the photoelectric heating of the IGM by a QSO is accomplished by soft
X--rays. The timescale required for the radiation near the light front to heat
the intergalactic gas to a temperature above that of the CMB is typically
10\% of the Hubble time. The \HII region produced by a QSO will therefore be
preceded by a warming front. 

An additional heating source is the \Lya photon scattering itself (MMR). 
The heating rate by the recoil of scattered \Lya photons is
\begin{equation}
\dot E_\alpha=\left(\frac{h_P\nu_{\alpha}}{m_\nH c^2}\right)h_P\nu_{\alpha}
P_\alpha, \label{eq:edot}
\end{equation}
where $P_\alpha$ is the \Lya scattering rate per hydrogen atom, and $m_\nH$ is
the mass of a hydrogen atom. In the case of
excitation at the thermalization rate $P_{\rm th}$, equation~(\ref{eq:edot})
becomes
\begin{equation}
\dot E_{\rm th}={27 h_P^2\nu_\alpha^2 A_{10}T_{\rm CMB}\over 4 m_\nH c^2 T_*} 
\approx
(315~{\rm K\,Gyr}^{-1})\left({{1+z}\over10}\right). \label{eq:Ethdot}
\end{equation}
The characteristic timescale for heating the medium above the CMB temperature
via \Lya resonant scattering at this rate is 
\begin{equation}
\Delta t={2 m_\nH c^2\nu_{10}\over 9 h_P\nu_\alpha^2}
A_{10}^{-1}\approx 10^8\,{\rm yr}, \label{eq:theat}
\end{equation}
about 20\% of the Hubble time at $z=8$. This result has an important
consequence: if a substantial population of sources of UV radiation turn on 
quite rapidly at (say) $z\sim 10$, there exists a finite interval
of time during which \Lya photons couple the spin temperature to the kinetic
temperature of the IGM before heating the gas above the CMB (reionization
occurs at even later times).
A window is then created in redshift space where a large 
fraction of intergalactic gas may be observed at $\sim 160\,$MHz in {\it  
absorption} against the CMB. We will show next how this transient phase
may imprint a strong signature on the radio sky. 

\section{Radio signatures}

\subsection{The ``cosmic web'' in emission at 21--cm}

Numerical $N$--body/hydrodynamics simulations of
structure formation within the framework of CDM--dominated cosmologies
(Cen \etal 1994; Zhang \etal 1995; Hernquist \etal 1996; Zhang \etal 1998) have
recently provided a definite picture for the topology of the IGM at $z\lta 5$,
one of an interconnected network of sheets and filaments with
virialized systems located at their points of intersection. In between the
filaments are underdense regions. This ``cosmic web'' appears to be a generic 
feature of CDM (Pogosyan \etal 1998), a pattern imprinted in the 
matter fluctuations at decoupling and sharpened by gravity. In a UV 
photoionizing background, modestly overdense filaments ($1<\rho_b/{\bar \rho_b}
<5$) will give rise to the \Lya forest seen in the spectra of high redshift 
quasars. As discussed below, radio observations at meter wavelengths on 
several arcmin scales may probe the cosmic web at times prior to the epoch of 
reionization.

Because of the clumpiness of the IGM, the 21--cm optical depth of
equation~(\ref{eq:tau}) will vary for different lines of sight.
To estimate the {\it rms} level of brightness temperature fluctuations
at $21 (1+z)\,$ cm, we shall consider three different
cosmological models with parameters suggested by a variety of recent
observations: an open model (OCDM, $\Omega_M=0.4$, $h=0.65$, $n=1$), a flat
model ($\Lambda$CDM, $\Omega_M=0.3$, $\Omega_\Lambda=0.7$, $h=0.7$, $n=1$),
and an Einstein--de Sitter tilted model (tCDM, $\Omega_M=1$,
$\Omega_\Lambda=0$, $h=0.5$, $n=0.8$). In all cases the amplitude of the power
spectrum or, equivalently, the value of the {\it rms} mass fluctuation
in a $8 \; h^{-1}$ Mpc sphere, $\sigma_8$, has been fixed in order to
reproduce the observed abundance of rich galaxy clusters in the local
universe (e.g. Eke \etal 1998).  The tilted model has been designed to
also match {\it COBE} measurements of the CMB on large scales. Our
choice of an open CDM cosmology (with a normalization which is 2 {\it
rms} away from the value observed by {\it COBE}) is guided by the
growing evidence in favour of a low value of $\Omega_M$ from cluster
studies (e.g. Carlberg \etal 1997).  The
$\Lambda$--dominated model matches both the cluster constraint on small scales
and {\it COBE} constraint on large. It is also able to account for most known
observational constraints, ranging from globular clusters ages to CMB
anisotropies (Ostriker \& Steinhardt 1995) to recent analyses of the
Type Ia SNe Hubble diagram (Perlmutter \etal 1998; Riess \etal 1998),
and just fits within the constraints imposed by gravitational lensing
(Kochanek 1996; Cooray \etal 1999). In all cosmologies the baryon density is
$\Omega_bh^2=0.02\,$ (Burles \& Tytler 1998).  The model
parameters are sumarized in Table \ref{cosmo}.
\begin{deluxetable}{l l l l l l}
\footnotesize
\tablenum{1}
\tablecaption{Cosmological parameters.  \label{cosmo}}
\tablewidth{0 pt}
\tablehead{
\colhead{Model} & $\Omega_M$ & $\Omega_\Lambda$
& $h$ & $n$ & $\sigma_8$
}
\startdata
tCDM  &  $1.0$ & $0.0$ & $ 0.5$ & $0.8$ & $ 0.55 $\nl
OCDM  &  $0.4$ & $0.0$ & $ 0.65$ & $1.0 $ & $ 0.92 $ \nl
$\Lambda$CDM &$0.3$ & $0.7$ & $ 0.7$ & $1.0$ & $1.1$ \nl
\enddata
%\tablecomments{}
\end{deluxetable}

As an illustrative example, consider a scenario in which sources of
\Lya photons are in sufficient abundance throughout the universe to
preheat the IGM to a temperature well above that of the CMB, and to
couple the spin temperature to the kinetic temperature of the
intergalactic gas everywhere. While the average 21--cm signal is two
orders of magnitude lower than that of the CMB (and will be swamped by the
much stronger non--thermal backgrounds that dominate the radio sky at
meter wavelengths and that must be removed), its fluctuations relative 
to the mean of the surveyed area -- induced in this example by density 
inhomogeneities only -- will greatly exceed those of the CMB. Brightness 
temperature fluctuations will be present
both in frequency and in angle across the sky. At high redshifts, and with
a spatial resolution of $\sim 1$ arcmin, a patchwork of radio emission 
will be produced by
regions with overdensities still in the linear regime and by
voids. From equations~(\ref{eq:tau}) and (\ref{eq:dT}), the {\it rms}
temperature fluctuation relative to the mean, neglecting angular
correlations in the projected IGM density, is
\begin{equation}
\langle \delta T_b^2\rangle^{1/2}\approx (2.9\,{\rm mK})\, h^{-1} \, 
\Big({{\Omega_b\, h^2}\over{0.02}}\Big) {(1+z)^2 
\sigma_\rho\over [\Omega_M(1+z)^3+\Omega_K(1+z)^2+\Omega_{\Lambda}]
^{1/2}}, \label{eq:deltat}
\end{equation}
where $\Omega_K=1-\Omega_M-\Omega_{\Lambda}$ is the curvature
contribution to the present density parameter.  This corresponds 
to a {\sl rms} fluctuation in the observed flux of
\begin{equation}
\langle dI^2\rangle ^{1/2}= (15 \,{\rm \mu Jy~arcmin^{-2}})\, h^{-1}
\, \Big({{\Omega_bh^2}\over{0.02}}\Big) {\sigma_\rho\over 
[\Omega_M(1+z)^3+\Omega_K(1+z)^2+\Omega_{\Lambda}]^{1/2}}. \label{emission}
\end{equation}
Here $\sigma_\rho$ is the variance of the density field in a volume
corresponding to a given bandwidth [$\Delta\nu/\nu=\Delta z/ (1+z)$]
and angular size $\Delta\theta$. The former corresponds to a comoving
length $L\approx (1+z)c H(z)^{-1}\Delta\nu / \nu\,$ Mpc, where $H(z)$ is
the Hubble constant and $\nu=1.4/(1+z)$ GHz is the observation
frequency, the latter to a comoving radius $R=0.5 \Delta\theta (1+z)/d_a(z)$, 
where $d_a$ is the angular diameter distance.  

On the scales of interest here, 
one can use the linearly evolved power spectrum $P(k)$ (Bardeen \etal 1986)
and a filter function that is a cylinder of radius $R$ and height $L$, to 
compute the density variance:
\begin{equation}
\sigma_\rho={{8 D^2(z)}\over {\pi^2 R^2L^2}}\int^\infty_0 dk 
\int^1_0 dx {{\sin^2(kLx/2) J_1(kR\sqrt{1-x^2})}
\over {x^2(1-x^2)}} (1+fx^2)^2{{P(k)}\over{k^2}},
\end{equation}
where $J_1$ is the Bessel function of order one, $D(z)$ is the linear growth 
factor (e.g. Peebles 1993), and $f\approx \Omega_M^{0.6}$ takes into account 
the effect of the peculiar velocities in compressing the same emitting volume 
into a narrower bandwidth (as the Hubble flow is reduced in overdense regions,
see Kaiser 1987). In this regime it is also legitimate to assume that the
baryons follow the dark matter, since segregation effects are important only on
smaller scales. We have verified using $N$--body simulations the validity of
the linear approximation and the neglect of angular correlations on scales
of $\sim 1$ arcmin. 

Figure 1 shows the fluctuation levels at 150 MHz ($z=8.5$) as a function of 
beam size for the models listed in Table 1. At a fixed bandwidth of 1 MHz, 
$\langle \delta T_b^2\rangle^{1/2}$ increases with decreasing angular scale 
from about 1 to 10 mK as $\sigma_\rho$ increases with decreasing linear scale. 
Since the growth of density fluctuations ceases early on in an open universe
(and the power spectrum is normalized to the abundance of clusters today),
the signal at a given $\Delta\theta$ is much larger in OCDM than in tCDM at 
high redshift. In a $\Lambda$CDM universe the level of fluctuations is
comparable with OCDM. The range of density fluctuations that would be
detectable at the $5\sigma$ level by a projected facility like the {\it SKA}
is shown in Figure 2, as a function of beam size and
frequency bandwidth, and for several assumed integration times. The
detection thresholds are scaled according to a theoretical {\it rms} continuum
noise in a 80~MHz bandwidth at 150~MHz of 64~nJy over an 8 hour integration,
allowing for only 21 of the 30 dishes in a compact array configuration
(Taylor \& Braun 1999)\footnote{See also http://www.nfra.nl/skai/science/.},
\begin{equation}
rms=(0.23\, \mu{\rm Jy}) \Big( {{1\, \rm MHz}\over{\Delta \nu}}\Big)^{1/2}
\Big({{100\, {\rm hours}}\over {\tau}}\Big)^{1/2}.
\label{noise}
\end{equation}
The central compact array in the `straw man' {\it SKA} design proposal has a 
diameter of $\sim50$~km in order to achieve 1--10 arcsec resolution. A more
appropriate
resolution for measuring fluctuations in the IGM intensity is $\sim1$ arcmin.
The compact array would have then to be scaled down to
$D_c\approx7\, {\rm km}\, (\Delta\theta/{\rm arcmin})^{-1}$ to be optimized
for such an experiment.
The dashed lines show curves of constant {\it rms} fluctuation of the 
signal in the corresponding beam. Each angular resolution 
observation corresponds to a different
compact array diameter $D_c$ (in other words, the 
array is being optimally resized to match its resolution to the desired
angular scale). Because the received flux is proportional
to the solid angle of the beam, the signal decreases with decreasing
angular scale until it falls below the detection threshold, indicated
by the solid lines. With 100 hours of observation in a tCDM model, 2 $\mu$Jy
fluctuations will be detectable only on scales exceeding 3 arcmin.
In OCDM, and similarly in $\Lambda$CDM, a signal of $\sim 2\,\mu$Jy may be
detected at significantly higher resolution with the same integration time.  

Figure 3 show radio maps at $z=8.5$ for tCDM and OCDM, assuming a
beamwidth of 2 arcmin and a frequency window of 1 MHz around $150$
MHz. Here a collisionless $N$--body simulation with 64$^3$ particles has
been performed with Hydra (Couchman, Thomas, \& Pearce 1995). The simulation
box size is $20 h^{-1}$ comoving Mpc, corresponding to 17 (11) arcmin in
tCDM (OCDM). The baryons are assumed to trace the dark matter
distribution without any biasing.  From a visual inspection of the two
simulated images, as well as from Figure 2, it seems possible that
observations with a modified (much less spread out) {\it SKA} design may be
used to reconstruct the matter density
field at redshifts between the epoch probed by galaxy surveys and
recombination, on scales as small as $0.5-2$ $h^{-1}$
comoving Mpc, i.e. masses in the range between $10^{12}$ and $10^{13}\; h^{-1}
M_\odot$.

\subsection{Observing the epoch of the first stars}

In the real universe, the observed radio pattern on
the sky depend on a combination of the underlying Gaussian density
field and the distribution and luminosities of the sources that cause
the preheating and reionization of the IGM. In this and the following sections
we shall discuss the detectability of the first radiation sources
in the universe through their impact on the surrounding neutral IGM.

In hierarchical clustering theories for the origin of cosmic structure,
massive objects grow `bottom--up' -- low mass perturbations
collapse early, then merge into progressively larger systems. In such a
scenario, one naturally expects the high redshift universe to contain small
scale substructures, systems with a virial temperature well below $10^4\,$K.
The resultant virialized systems will remain as neutral gas clouds unless
they can
cool due to molecular hydrogen, in which case they may form Pop III stars.
These stars will produce a background of UV continuum photons near the \Lya frequency which 
will immediately escape into the IGM, while at the same time ionizing the gas in 
their vicinity. \Lya photons will propagate into uncollapsed, largely neutral regions 
of the IGM where the kinetic temperature in the absence of preheating will be
\begin{equation}
T_K(z)\approx 26\,{\rm mK}~(1+z)^2 \label{eq:Tc}
\end{equation}
(Couchman 1985), well below $T_{\rm CMB}$ because of adiabatic cooling
during cosmic expansion. If Pop III sources are in sufficient
abundance throughout the universe, the \Lya flux will couple the spin
temperature to the kinetic temperature, and $T_S$ will be pulled below
$T_{\rm CMB}$ everywhere. After about $10^8$ yr (see
eq. \ref{eq:theat}), the same \Lya photons responsible for the coupling
will warm the IGM to a temperature well above that of the CMB.
Eventually a sufficient number of photons above 1 Ryd may be able to
penetrate all of the surrounding neutral gas in the cloud, reach the
external intergalactic gas, and create expanding intergalactic \HII
regions.

The crucial assumption we make here is that the \Lya radiation field
switches on at a redshift $z>z_{\rm th}$ following the formation of
the first stars, and reaches the thermalization value $P_{\rm th}$ at
$z_{\rm th}$ on a timescale much shorter than the Hubble time at that
epoch.  While in hierarchical models of structure formation the spatial number
density of halos of a given mass grows initially exponentially (Press
\& Schechter 1974), the rise time of the UV radiation background will
be modulated by gasdynamical processes. In the model of Gnedin \&
Ostriker (1997), for instance, a \Lya metagalactic flux is rapidly
produced in the range $7<z<10$.  In this case
coupling between $T_S$ and $T_K\ll T_{\rm CMB}$ is almost
instantaneous at $z=z_{\rm th}$, and the IGM is suddenly detectable in
absorption. For $z<z_{\rm th}$, \Lya photons begin to
heat the IGM and the kinetic temperature increases at the rate
\begin{equation}
{dT_K\over dz}= {{2\mu}\over 3} 
{\dot E_{\rm th}\over k_B} {{dt}\over{dz}} + 2{{T_K}\over{(1+z)}},
\label{tk}
\end{equation}
until $T_K>T_{\rm CMB}$ and the IGM is now detectable in emission.
Here, $\mu=16/13$ is the mean molecular weight for a neutral gas with a
fractional abundance by mass of hydrogen equal to 0.75. Such a
chain of events will leave a strong imprint on the CMB, and mark the epoch 
of reheating by the first generation of stars. At 150 MHz ($z_{\rm th}=8.5$), 
the brightness temperature (shown in Figure 4) shows an absorption feature 
of 40 mK, with a width of about 10 MHz.
Such a depression in the CMB is much easier to observe compared with
brightness fluctuations, as the isotropic signal may be recorded with
a larger beam area.  Assuming $\Delta T\simeq 40$ mK as a fiducial
estimate, the signal--to--noise ratio for the {\it SKA} is
\begin{equation}
\frac{S}{N}\approx 9~ \Big( {{\Delta \nu}\over{1\, \rm MHz}}\Big)^{1/2}
\Big({\tau\over 100 \, {\rm hours}}\Big)^{1/2} \, A_b,
\label{noise2}
\end{equation} 
where $\Delta \nu$ is the frequency resolution, $A_b$ is the beam area
in arcmin$^2$, and $\tau$ is the observation time. While the strength of 
this feature depends mainly on the timescale $t_{\rm th}$ over which the 
continuum Ly$\alpha$ background field reaches the thermalization value, 
we have checked that the absorption dip is well above the detection limit 
for all timescales $t_{\rm th}\lta 0.1 t_H$, where $t_H$ is the then Hubble
time. Note that similar sharp signatures in the radio background at meter
wavelengths could also be produced at the reionization epoch, as recently
discussed by Gnedin \& Ostriker (1997) and Shaver \etal (1999). 

\subsection{Observing the epoch of the first quasars}

Very little is actually known observationally about the nature of the first 
bound objects and the thermal state of the universe at early epochs. 
If ionizing sources are uniformly distributed, like an abundant population of 
pregalactic stars, the ionization and thermal state of the IGM will be the
same everywhere at any given epoch, with the neutral fraction decreasing
rapidly with cosmic time. As discussed above, this may be the
case in CDM dominated cosmologies, since bound objects sufficiently
massive ($\sim 10^6 \msun$) to make stars form at high redshift. On the
other hand, reionization may also occur in a highly inhomogeneous fashion, 
as widely separated but very luminous
sources of photoionizing radiation such as QSOs, present at the time
the IGM is largely neutral, generate expanding \HII regions on Mpc scales:
the universe will be divided into an ionized phase whose filling factor
increases with time, and an ever shrinking neutral phase. If the ionizing 
sources are randomly distributed, the \HII regions will be spatially isolated
at early epochs. 

In such a scenario, 21--cm emission on Mpc scales will be produced in
the quasar neighborhood as the medium surrounding it is heated by soft
X--rays from the QSO itself (MMR).  Figure \ref{fig6} shows the (differential) 
radio map resulting
from a QSO `sphere of influence' soon after it turns on at $z=8.5$ (tCDM).
The kinetic temperature profile and radiation field around the quasar were
computed including finite light travel time effects and heating by secondary
electrons collisionally produced by primary electrons photo--ejected by the
soft X--rays (MMR). The specific luminosity of the QSO was taken to be
proportional to $\nu^{-1.8}$, normalized to a production rate of \HI ionizing
photons of $10^{57}\, {\rm s}^{-1}$.  The IGM was assumed uniform with
$\Omega_b=0.08$ and $H_0=50\kmsmpc$. The resulting radial temperature profiles
were then superimposed on the surrounding density fluctuations as computed
using Hydra. The visual effect is due to the convolution of the spin
temperature profile with the (linearly) perturbed density field around the
quasar. Outside the \HII bubble, there is an inner
thin shell of neutral gas where the IGM is heated to $T_S=T_K>T_{\rm CMB}$
as the hyperfine levels are mixed by \Lya continuum photons from the QSO, and
a much larger external shell where $T_S=T_K<T_{\rm CMB}$ because of adiabatic
cooling. At larger distances from the quasar the \Lya coupling
strength is weakening as $r^{-2}$, and $T_S\rightarrow T_{\rm CMB}$. As
the warming front produced by the quasar expands, a growing amount of
the surrounding IGM is unveiled both in emission and in
absorption. The size and intensity of the detectable 21--cm region 
depend on the QSO luminosity and age. In the figure the signal ranges 
from $-3 \mu$Jy to $3 \mu$Jy per beam.

Note that, although the quasar was placed in a high density
region in the corner of the simulation volume, the figure can equally
be viewed as the emission due to heating by a beam of soft X--rays from
the quasar with an opening angle of $90^\circ$.  Thus imaging the gas
surrounding a quasar in 21--cm emission could provide a direct means of
measuring the opening angle of quasar emission. 

\section{A compact, special purpose, low frequency array}

The requirements for observing the epoch of IGM reheating and reionization
are a narrow bandwidth ($\Delta\nu\approx$ 1 MHz), to achieve adequate
resolution in velocity space at the relevant redshift, and an angular resolution
of a few arcminutes, the typical scale of the expected density fluctuations.
The straw man proposal for the {\it SKA} suggests a large diameter for its 
central
compact array in order to achieve a resolution of $\sim 1-10$ arcsec. While
gas density fluctuations would persist to smaller scales in the IGM, Figure 2
shows that the associated signal would drop well below the threshold for a
practical detection. For a successful experiment it is necessary to 
significantly decrease the diameter of the compact array. The current design
for the {\it SKA} (and the {\it GMRT}) has too widely distributed 
`sub--apertures' for a program oriented toward detecting arcmin scale
structures.

An alternative possibility is to build a compact, special purpose radio 
telescope dedicated to searching for the expected reheating and
reionization signals at low frequencies. A single dish with a diameter of 
200~m (smaller than Arecibo) would resolve scales of 30 arcminutes, 
corresponding to $35-50 h^{-1}$ 
comoving Mpc. This is comparable to the largest scale expected for coherent 
primordial density
fluctuations (Pogosyan \etal 1998). Figure 1 shows that the expected brightness
temperature fluctuations on this scale will be in excess of 1~mK. The {\it rms}
brightness temperature is given by the radiometer equation as
\begin{equation}
\Delta T_B=(0.33\, {\rm mK}) \left(\frac{T_{\rm sys}}{200\, {\rm K}}\right)
\left( {{1\, \rm MHz}\over{\Delta \nu}}\right)^{1/2}
\left({{100\, {\rm hours}}\over {\tau}}\right)^{1/2}
\label{eqn:radiometer}
\end{equation}
(Burke \& Graham--Smith 1997) for a bandwidth $\Delta\nu$ and integration
time $\tau$. The system temperature $T_{\rm sys}$ is sky--limited to 200~K
in the coldest directions through the Galaxy (Burke \& Graham--Smith).
If a density fluctuation
filled the beam, it would be detectable at the $5\sigma$ level in a 300 hour
integration.

Since such structures tend to be filamentary, a more typical covering
factor may be $\sim0.3$, increasing the required integration time by a
factor of 10. This may be offset by increasing the bandwidth. The velocity
width of a $50 h^{-1}$ comoving Mpc filament at $z=8.5$ corresponds to a
frequency width at 150~MHz of about 8~MHz. A bandwidth of 5~MHz would then
require an integration time of 500 hours. Such a survey operating over several
years would enable a substantial range in redshift to be probed.

\section{Conclusions}

The history of the universe between the epoch of recombination and the
birth of bright sources of radiation by $z\lta5$ is currently shrowded in
mystery:\ the absence of radiation sources leaves this period a ``dark age.''
Yet spectra of QSOs show not only metal absorption features in their spectra,
the signature of ongoing and past star formation, but evidence for non--linear 
density fluctuations on large scales as well, a ``cosmic web'' in the form of 
the \Lya forest.

We have shown that it may be possible to detect both the development of the
cosmic web and the epoch of the first radiation sources, whether stars or QSOs,
using 21--cm measurements from intergalactic gas. Prior to reionization, the
radio signal will display emission and/or absorption structures both
in angle across the sky and in redshift. Fluctuations in 21--cm emission will
result from non--uniformities in the gas density and from non--uniformities in
the distribution of \Lya radiation sources required to decouple the spin
temperature from the CMB temperature. With an instrument like the
{\it Square Kilometre Array}, such fluctuations would be detectable on arminute
scales, and so reveal whether the reheating is due to galaxies or QSOs.

Absorption measurements may permit the epoch of the first stellar sources to be
determined. The continuum radiation from an early generation of stars,
redshifted to \Lya, will couple the spin temperature to the low kinetic
temperature of the IGM for a transitory interval before heating the
gas above the CMB temperature. The result is a detectable absorption signature
against the CMB that would flag the first epoch of massive star
formation in the universe. If reheating were due to ultraluminous but
sparsely distributed sources (e.g., the first bright QSOs), the
resultant patches would have a larger scale than any
gravitationally--induced inhomogeneities. Because of the very large scales
involved, there is no difficulty in principle in achieving a sufficiently
narrow bandwidth and an adequate angular resolution with the {\it SKA} or
{\it GMRT}; the main limitations are the low flux levels and the radio
background subtraction. Alternatively, a special purpose telescope dedicated
to scanning the sky in frequency along cold lines of sight through the Galaxy
may be used to detect the epochs of reheating. Subsequent
high resolution measurements with the {\it SKA} would provide a means of
identifying the nature of the first radiation sources and so shedding 
new light on the thermal and reionization history of the universe.

\acknowledgements

Support for this work was provided by NASA through ATP grant NAG5--4236
(PM, AM, and PT), and by the Royal Society (MJR). We thank the anonymous
referee for clarifying the technical requirements needed for this
project. The results in this paper made use of the Hydra $N$--body code 
(Couchman, Thomas \& Pearce 1995).

\appendix
\section{The Wouthuysen--Field mechanism}

Electric dipole radiation (\Lya photons) can induce a spin transition through
spin--orbit coupling. The relevant quantum number is the total angular 
momentum $F=I+J$, where $I$ is the proton spin and $J$ is the total angular 
momentum of the electron, $J=S+L$. There are six hyperfine states in total 
that contribute to the \Lya transition. Only four of these, 
the $n=1$ singlet $_0S_{1/2}$ and triplet
$_1S_{1/2}$ states (the notation is $_FL_J$), and the two triplet $n=2$ states
$_1P_{1/2}$ and $_1P_{3/2}$ contribute to the excitation
of the 21--cm line by \Lya scattering. The selection rule $\Delta F=0,1$ 
permits the
transitions $_0S_{1/2}\rightarrow\, _1P_{1/2},\, _1P_{3/2}$ and
$_1P_{1/2},\, _1P_{3/2}\rightarrow\, _1S_{1/2}$, and so effectively
$_0S_{1/2}\rightarrow\, _1S_{1/2}$ occurs via one of the $n=2$ states.
Denoting the occupation number
of $_0S_{1/2}$ by $n_0$ and that of $_1S_{1/2}$ by $n_1$, the rate
equation for $n_0$ is
\begin{equation}
\frac{dn_0}{dt}=A_{10}\left(1+\frac{T_{\rm R}}{T_*}\right)n_1-
3A_{10}\frac{T_{\rm R}}{T_*}n_0
+P^\alpha_{10}n_1 - P^\alpha_{01}n_0, \label{eq:rate}
\end{equation}
where the radiation intensity at the 21--cm frequency has been expressed
in terms of the antenna temperature $T_{\rm R}$, $I_{10}=2k_{\rm B}T_{\rm R}/
\lambda_{10}^2$. The ratio $T_{\rm R}/T_*$ is the number of 21--cm photons
per mode. Here, $P^\alpha_{01}$ and $P^\alpha_{10}$ are the effective
excitation and de--excitation rates due to \Lya scattering. These may
be related to the total scattering rate of \Lya photons, $P_\alpha$,
as shown below.

It is convenient to label the $n=2$ levels as states $2-5$, from lowest energy
to highest. Then the external radiation intensity at the frequency $\nu_{ij}$
corresponding to the $i\rightarrow j$ transition may be expressed as the
antenna temperature $T_{\rm R}^{ij}$, as above. The temperature corresponding
to the energy difference itself, $h\nu_{ij}$, is expressed as $T_{ij}=
h\nu_{ij}/k_{\rm B}$. If $A_{ji}$ denotes the spontaneous decay rate
for the transition $j\rightarrow i$, then $P^\alpha_{01}$ and $P^\alpha_{10}$
are given by
\begin{equation}
P^\alpha_{01}=3\frac{T^{\rm R}_{03}}
{T_{03}}\frac{A_{30}A_{31}}{A_{30}+A_{31}}+3\frac{T^{\rm R}_{04}}
{T_{04}}\frac{A_{40}A_{41}}{A_{40}+A_{41}}
\end{equation}
and
\begin{equation}
P^\alpha_{10}=\frac{T^{\rm R}_{13}}{T_{13}}\frac{A_{30}A_{31}}{A_{30}+A_{31}}+
\frac{T^{\rm R}_{14}}{T_{14}}\frac{A_{40}A_{41}}{A_{40}+A_{41}}.
\end{equation}

The ratios $A_{ji}/A_\alpha$, where $A_\alpha$ is the total \Lya
spontaneous decay rate, may be solved for using a sum rule for the
transitions. This states that the sum of all transitions of given $nFJ$
to all the $n'J'$ levels (summed over $F'$ and $m_{F'}$) for a given
$n'J'$ is proportional to $2F+1$ (e.g. Bethe \& Salpeter 1957).
There are
four sets of downward transitions to $n'=1$, $J'=1/2$, corresponding
to the decay intensities $I_{51}$, $I_{50} (=0)$, $I_{41}$, $I_{40}$,
$I_{31}$, $I_{30}$, $I_{21}$, and  $I_{20} (=0)$. These give
\begin{equation}
\frac{I_{51}}{I_{40}+I_{41}}=\frac{5}{3},\quad
\frac{I_{40}+I_{41}}{I_{30}+I_{31}}=1,\quad
\frac{I_{30}+I_{31}}{I_{21}}=3.
\end{equation}
Similarly, there are four sets of upward transitions to $n'=2$,
$J'=1/2$ or $3/2$, giving the intensities $I_{30}$, $I_{20} (=0)$, $I_{31}$,
$I_{21}$, $I_{50} (=0)$, $I_{40}$, $I_{51}$, and  $I_{41}$. These ratios are
\begin{equation}
\frac{I_{40}}{I_{41}+I_{51}}=\frac{1}{3},\quad
\frac{I_{30}}{I_{21}+I_{31}}=\frac{1}{3}.
\end{equation}
Using $I_{kj}/I_\alpha=(g_k/g_{\rm tot})(A_{kj}/A_\alpha)$, where $I_\alpha$ is
the total \Lya decay intensity summed over all the hyperfine transitions, and
$g_k=2F+1$ is the statistical weight of level $k$ ($g_{\rm tot}$ is the sum
of the weights of the upper levels), gives $A_{20}/A_\alpha=A_{50}/A_\alpha=0$,
$A_{21}/A_\alpha=A_{51}/A_\alpha=1$, $A_{30}/A_\alpha=A_{41}/A_\alpha=1/3$,
and $A_{31}/A_\alpha=A_{40}/A_\alpha=2/3$. We then obtain $P_\alpha=3A_\alpha
(T_{\rm R}/T_\alpha)$ ($T_\alpha\equiv h\nu_\alpha/k_{\rm B}$), and
$P_{10}=(4/27)P_\alpha$. Statistical equilibrium for $T_{\rm R}=0$ requires
$P_{01}/P_{10}=n_1/n_0=3\exp(-T_*/T_K)$. Solving the rate equation~(\ref
{eq:rate}) for $T_{\rm R}>0$ gives equation~(\ref{eq:Tspin}), where $T_{\rm R}=
T_{\rm CMB}$, and where $T_S\gg T_*$, $T_K\gg T_*$, and $T_{\rm R}\gg T_*$ 
have been assumed.

\newpage

\begin{figure}
\plotone{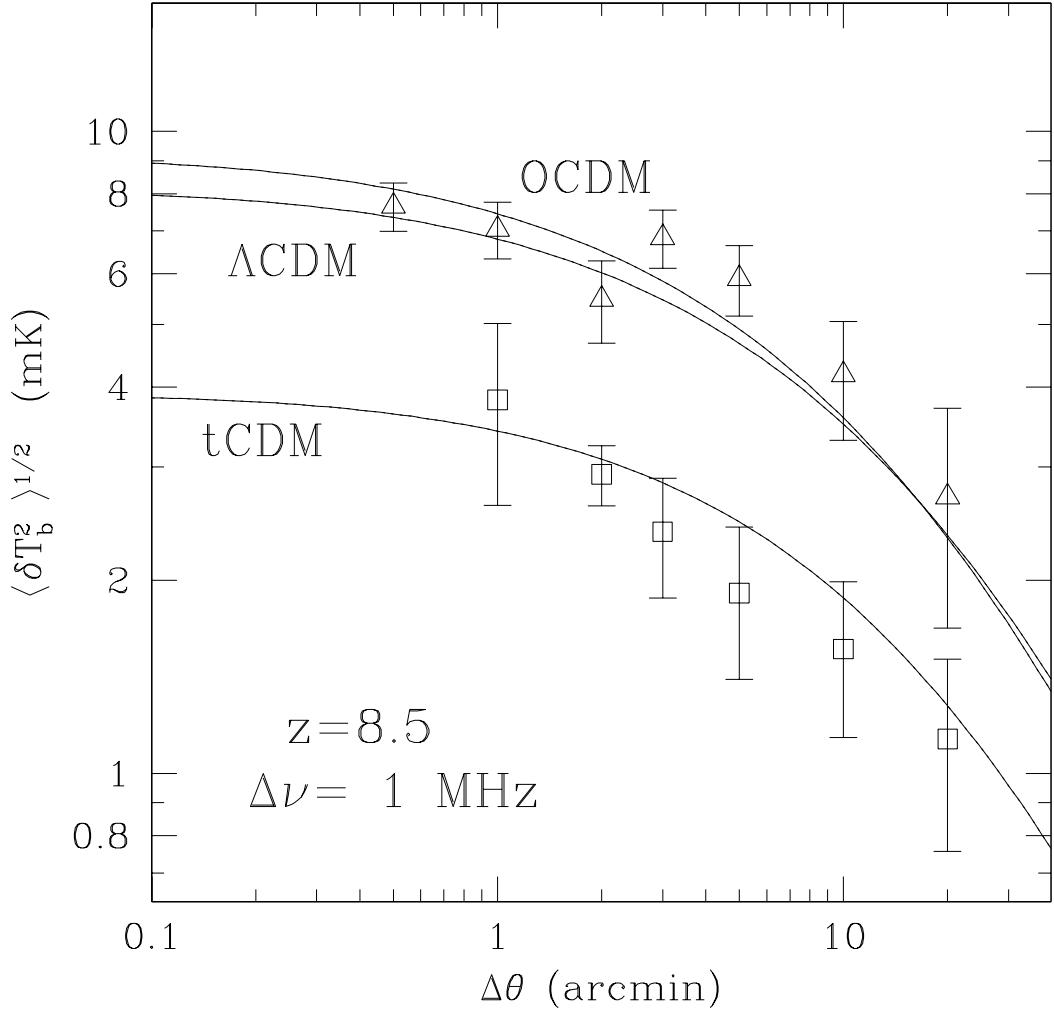}
\caption{Predicted {\sl rms} brightness temperature fluctuations
relative to the average at 150 MHz for the different cosmologies described
in the text, as a function of beam size.  A neutral medium with
spin temperature $T_S\gg T_{\rm CMB}$ has been assumed. The
bandwidth is 1 MHz (corresponding to a comoving length of 6.5 $h^{-1}$
Mpc for $\Omega_M=1$).  The data points with error bars show the
range of values found in a set of $N$--body simulations for tCDM ({\it 
squares}) and OCDM ({\it triangles}).
\label{fig1}}
\end{figure}

\begin{figure}
\vspace{9cm}
\plotone{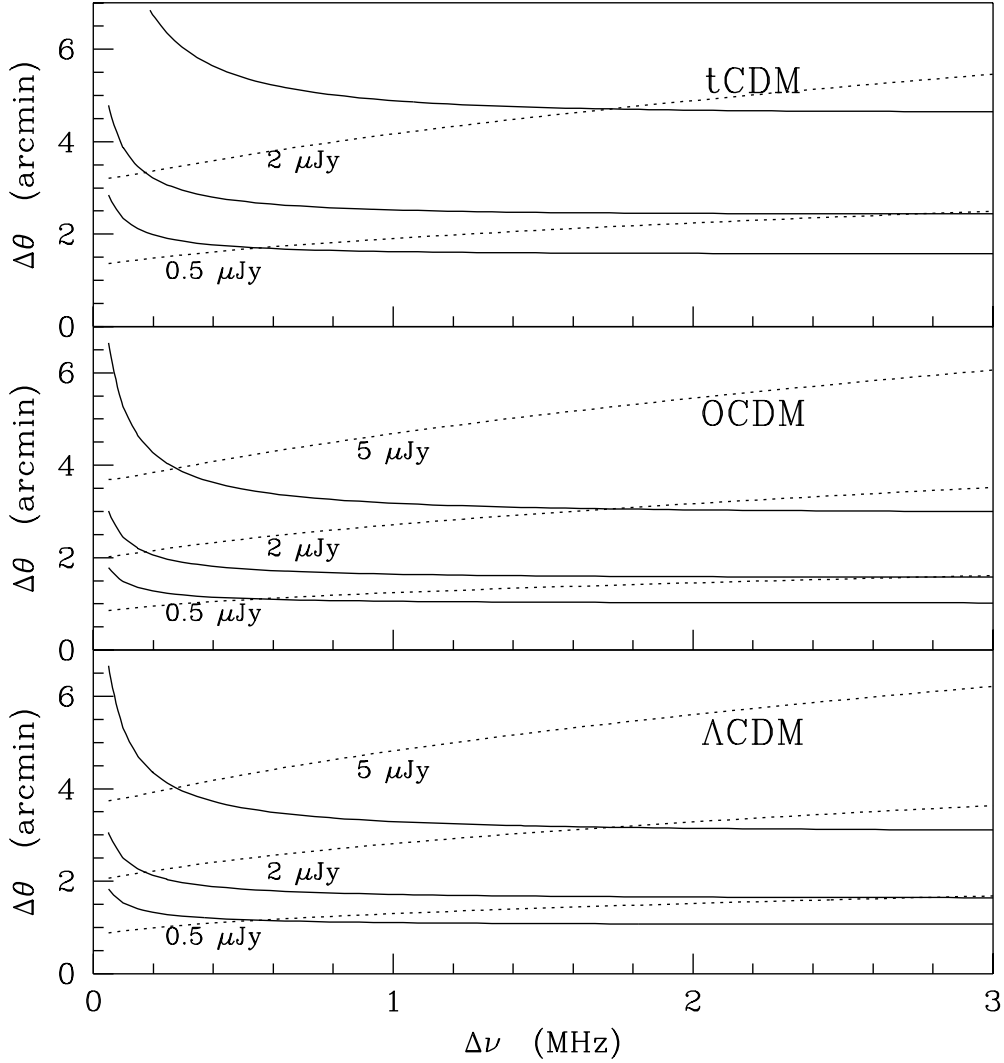}
\caption{Detectability of IGM fluctuations in 21--cm flux density per
beam (in $\mu$Jy) at 150 MHz ($z=8.5$) by a projected facility like the 
{\it Square Kilometer Array} as a function of 
beam width and bandwidth.  {\it Dashed lines:} Size of the {\it rms} 
fluctuations ($0.5\mu{\rm Jy}$, $2\mu{\rm Jy}$, and $5\mu{\rm Jy}$),
relative to the mean expected in a beam defined by the angular diameter
$\Delta\theta$ and bandwidth $\Delta\nu$.
{\it Solid lines:} Threshold below which the fluctuations are undetectable
($S/N<5$) 
with the {\it SKA} for integration times of 10, 100, and 500 hours (from 
top to bottom). Note that this calculation assumes that the array is   
optimally resized (becoming more compact at coarser angular resolution) 
in order to match its resolution to the desired angular scale.  
\label{fig2}}
\end{figure}

\newpage

\begin{figure}
\hspace{1.5cm}\plottwo{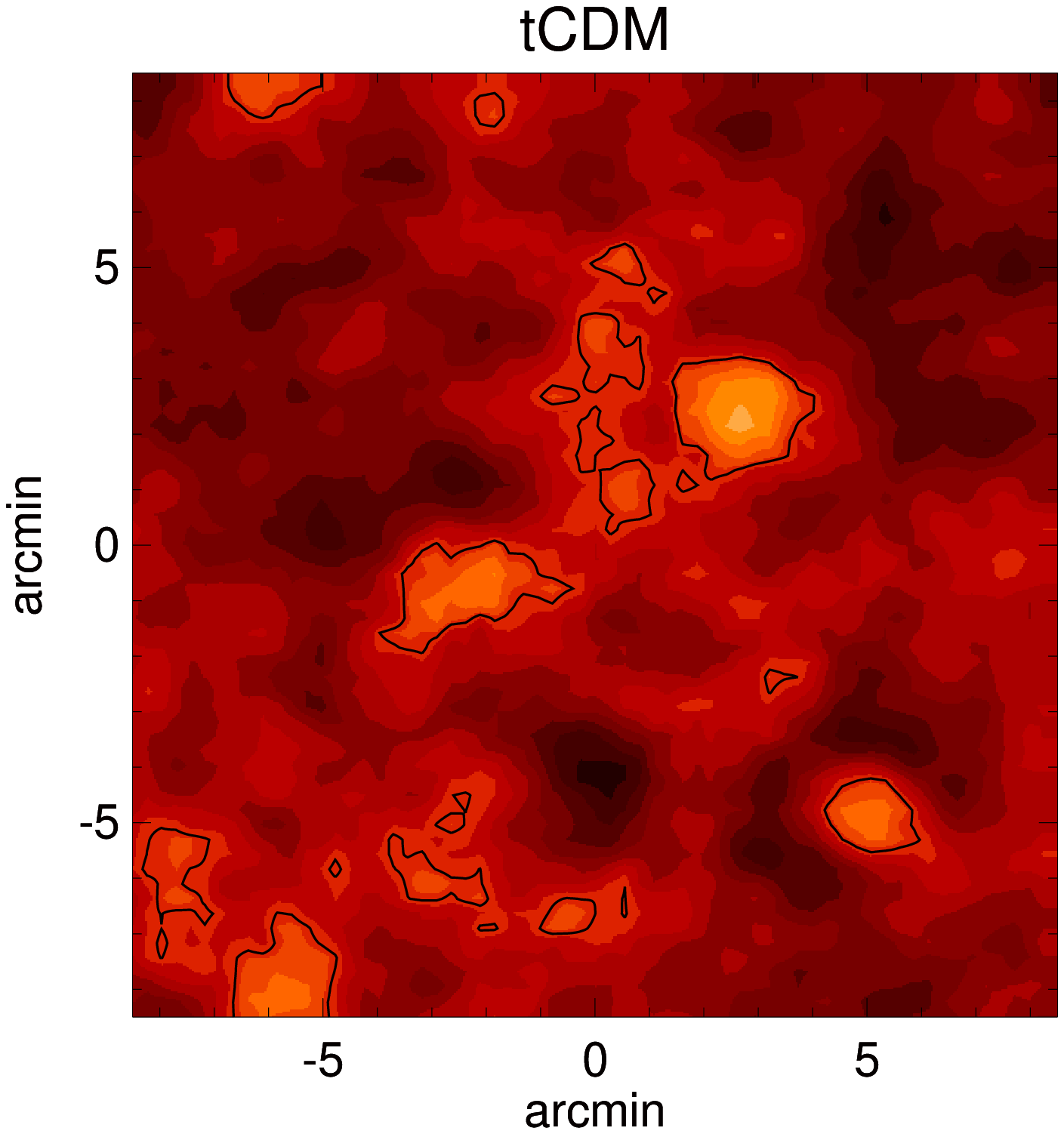}{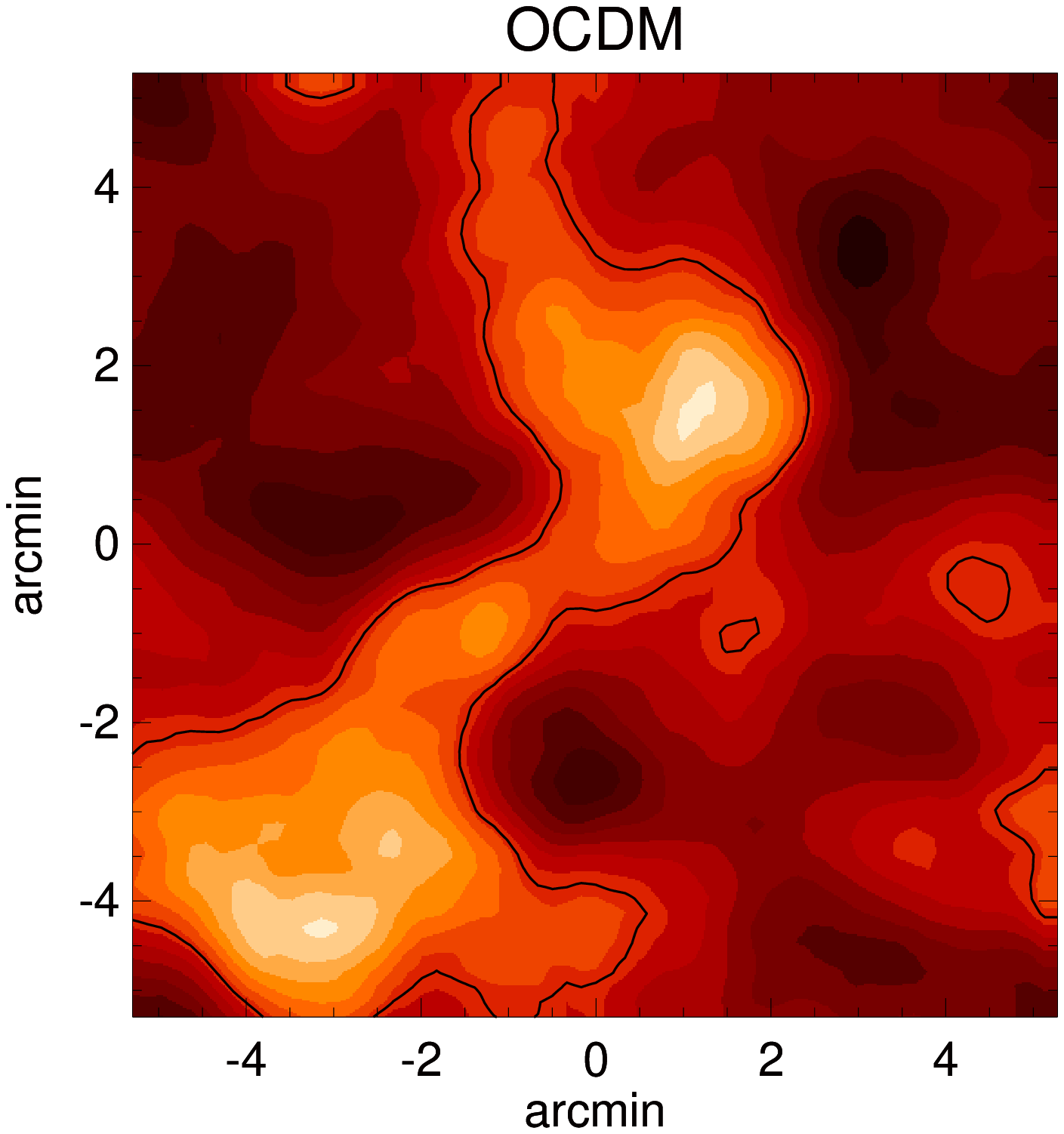}
\vspace{9cm}
\caption{{\it Left:} Radio map of redshifted 21--cm emission against the CMB
in a tCDM cosmology at $z=8.5$.  The linear size of the box is $20h^{-1}$ 
(comoving) Mpc. The point spread function of the synthesized beam 
is assumed to be a spherical top--hat with a width of 2 arcmin. The 
frequency window is 1 MHz around a central frequency of $150$ MHz.
The color intensity goes from $1$ to $6\,$ $\mu$Jy per beam.
For clarity, the contour levels outline regions with
signal greater than $4\,\mu$Jy per beam. {\it Right:} Same for OCDM.
\label{fig3}}
\end{figure}

\begin{figure}
\vspace{9cm}
\plotone{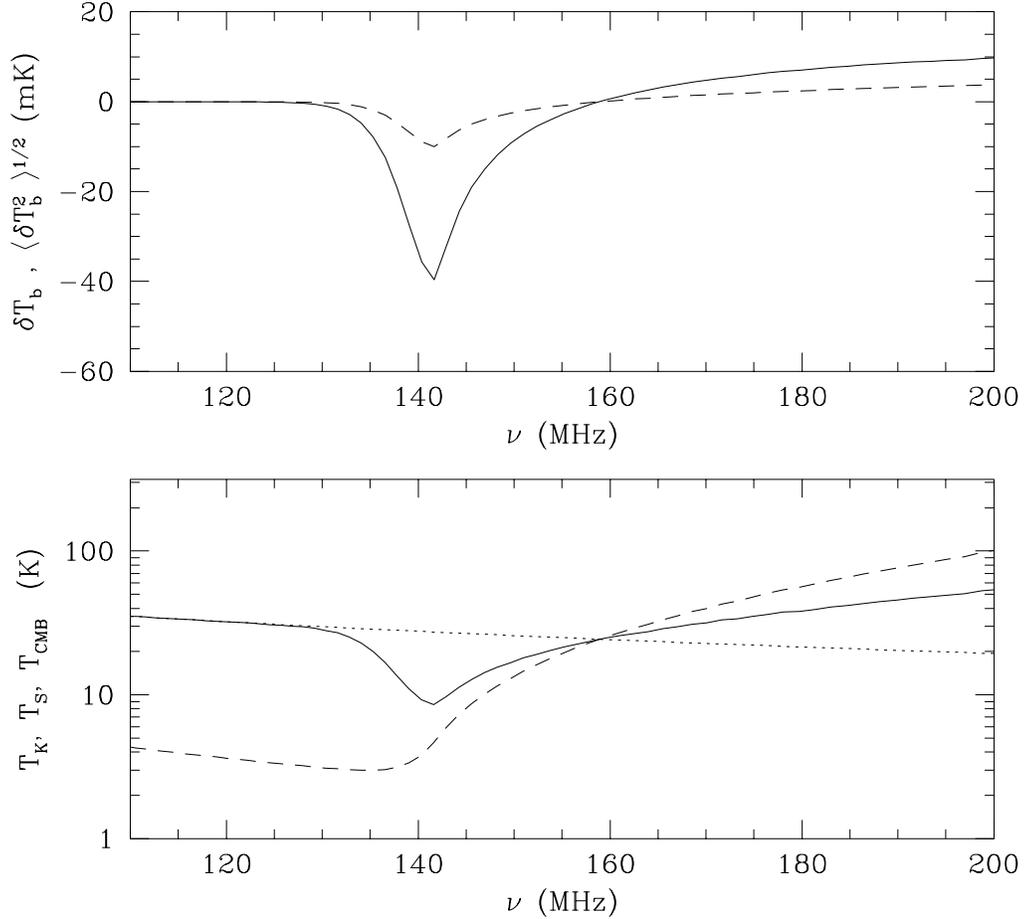}
\caption{{\it Top:} Mean $\Delta T=T_b-T_{\rm CMB}$ versus frequency
for an angular resolution of 1 arcmin and frequency resolution of 1
MHz ({\it solid line}), together with the {\sl rms} fluctuation value
({\it dashed line}). The IGM is reheated at $z_{\rm th}=9$.  The
strong absorption feature in the mean antenna temperature is associated 
with the fast rise of a \Lya continuum background on a time scale $\approx 
10$ Myr (see text),  coupling $T_S$ with $T_K$.  {\it Bottom:} Corresponding 
evolution of the kinetic ({\it dashed line}), spin ({\it solid line}), and 
CMB temperatures ({\it dotted line}).
\label{fig4}}
\end{figure}

\begin{figure}
\hspace{1.5cm}\plottwo{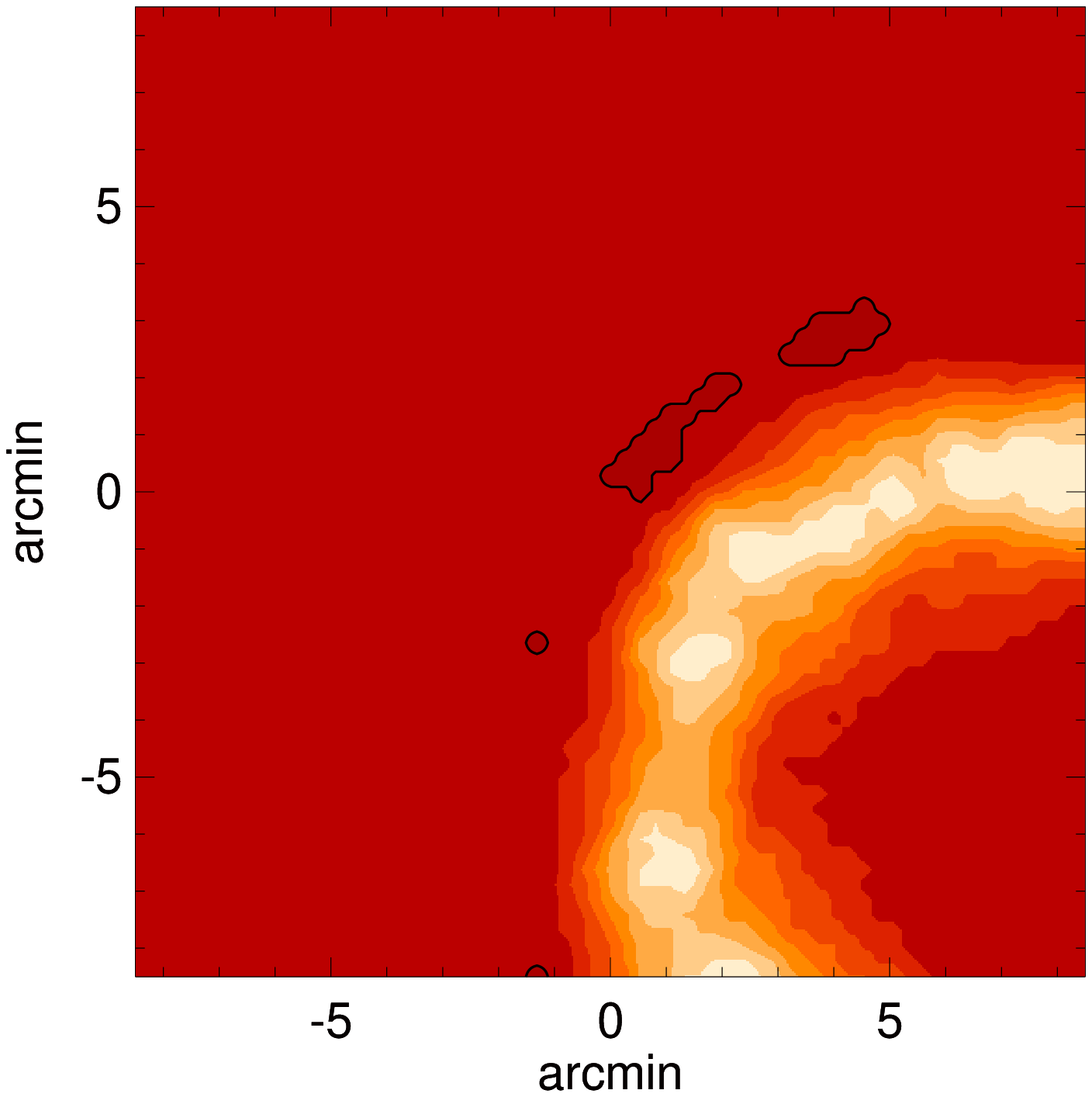}{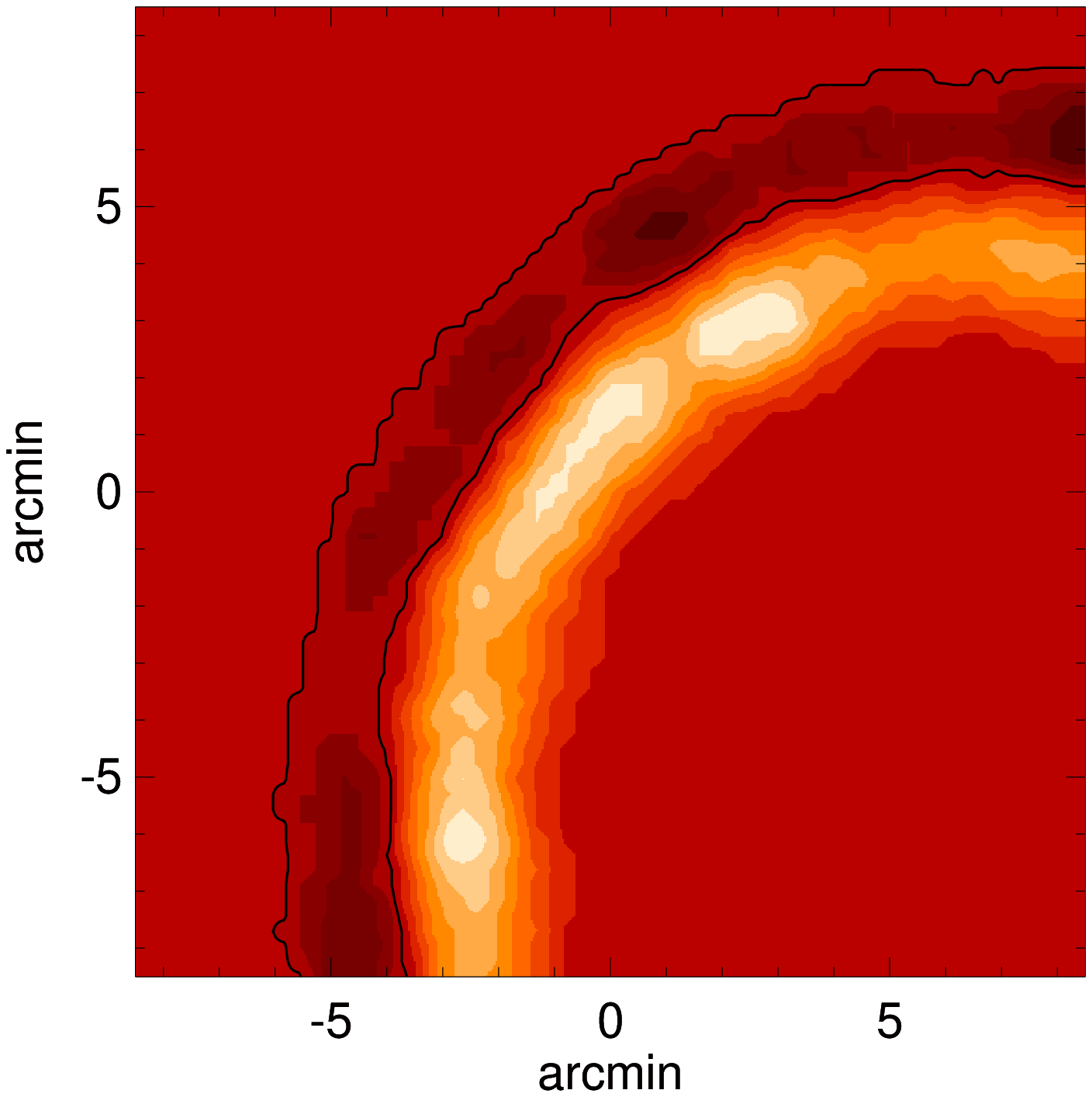}
\vspace{9cm}
\caption{21--cm emission and absorption against the CMB from the region 
surrounding a quasar source
(lower right corner, in the center of an \HII zone), revealed once
the IGM is heated above the CMB by soft X-rays from the quasar.
The linear size of the box is of $20h^{-1}$ comoving Mpc (tCDM), the angular 
resolution is $2$ arcmin, and the bandwidth is 1~MHz. The color levels
range from $-3$ $\mu$Jy to $3$ $\mu$Jy per beam. The dark contour demarcates
the absorption ring. The quasar turns on at $z = 8.5$ with an ionizing photon
luminosity of $10^{57}$ photons s$^{-1}$, and is
observed after $7$ ({\it left}) and $10$ ({\it right}) Myr.  The
temperature of the IGM beyond the light radius is assumed to be
$T_K=26\, {\rm mK}\, (1+z)^2\approx 2.3$ K $<T_{\rm CMB}$.
\label{fig6}}
\end{figure}

\end{document}